\documentclass[journal,twoside,web]{ieeecolor}

\usepackage{bm}
\usepackage{generic}
\usepackage{cite}
\usepackage{amsmath,amssymb,amsfonts}
\usepackage{graphicx}
\usepackage{textcomp}

\usepackage{algorithm,algorithmic}
\usepackage{float}
\usepackage{epsfig}
\usepackage{graphics}
\usepackage{multicol}
\usepackage{lipsum}
\usepackage{subfigure}
\usepackage{url}
\usepackage{verbatim}
\usepackage{xspace}
\floatstyle{ruled}
\newfloat{model}{H}{mod}
\floatname{model}{\footnotesize Model}
\newfloat{notatio}{H}{not}
\floatname{notatio}{\footnotesize Notation}

\newenvironment{varalgorithm}[1]
  {\algorithm\renewcommand{\thealgorithm}{#1}}
  {\endalgorithm}

\newenvironment{list4}{
	\begin{list}{$\bullet$}{%
			\setlength{\itemsep}{0.05cm}
			\setlength{\labelsep}{0.2cm}
			\setlength{\labelwidth}{0.3cm}
			\setlength{\parsep}{0in} 
			\setlength{\parskip}{0in}
			\setlength{\topsep}{0in} 
			\setlength{\partopsep}{0in}
			\setlength{\leftmargin}{0.16in}}}
	{\end{list}}

\newenvironment{list4a}{
	\begin{list}{$\bullet$}{%
			\setlength{\itemsep}{0.05cm}
			\setlength{\labelsep}{0.2cm}
			\setlength{\labelwidth}{0.3cm}
			\setlength{\parsep}{0in} 
			\setlength{\parskip}{0in}
			\setlength{\topsep}{0in} 
			\setlength{\partopsep}{0in}
			\setlength{\leftmargin}{0.16in}}}
	{\end{list}}

\newcommand{\acronym}{{$\mathrm{DMaC}$}\xspace}
\newcommand{\acronymbf}{{$\mathbf{DMaC}$}\xspace}

\def\BibTeX{{\rm B\kern-.05em{\sc i\kern-.025em b}\kern-.08em
    T\kern-.1667em\lower.7ex\hbox{E}\kern-.125emX}}
\usepackage{cite}
\usepackage{verbatim}
\usepackage{hyperref}
\usepackage{soul}
\hypersetup{
     colorlinks = true,
     linkcolor = [rgb]{0,0,1},
     anchorcolor = [rgb]{0,0,1},
     citecolor = [rgb]{0.9,0.5,0},
     filecolor = [rgb]{0,0,1},
     urlcolor = [rgb]{0.9,0.5,0},
        citecolor = [rgb]{0.9,0.5,0},
        anchorcolor = [rgb]{0.2,0.6,0.2},
        hyperindex = true,
        hyperfigures
}
\newtheorem{theorem}{\bfseries Theorem}

\newtheorem{prop}{\bfseries Proposition}

\newtheorem{assum}{\bfseries Assumption}

\newtheorem{example}{\bfseries Example}
\newtheorem{remark}{\bfseries Remark}
\begin{document}

\title{Max-Consensus with Deterministic Convergence in Directed Graphs \\ with Unreliable Communication Links}

\author{Apostolos I. Rikos, \IEEEmembership{Member, IEEE}, Jiaqi Hu, Themistoklis Charalambous, \IEEEmembership{Senior Member, IEEE}, \\ and Karl Henrik Johannson, \IEEEmembership{Fellow, IEEE}
\thanks{Apostolos~I.~Rikos and Jiaqi~Hu are with the AI Thrust, Information Hub, The Hong Kong University of Science and Technology (Guangzhou), Guangzhou, China. 
Apostolos~I.~Rikos is also affiliated with the Department of Computer Science and Engineering, The Hong Kong University of Science and Technology, Clear Water Bay, Hong Kong. 
E-mails: {\tt~apostolosr@hkust-gz.edu.cn; jhu021@connect.hkust-gz.edu.cn}}
\thanks{T.~Charalambous is with the Department of Electrical and Computer Engineering, University of Cyprus, 1678 Nicosia, Cyprus. He is also a Visiting Professor with the Department of Electrical Engineering and Automation, School of Electrical Engineering, Aalto University. E-mail:{\tt~charalambous.themistoklis@ucy.ac.cy}.}
\thanks{K.~H.~Johansson is with the Division of Decision and Control Systems, KTH Royal Institute of Technology, SE-100 44 Stockholm, Sweden. He is also affiliated with Digital Futures. E-mail:{\tt~kallej@kth.se}.}
\thanks{This work was supported by the Guangzhou-HKUST(GZ) Joint Funding Scheme (Grant No. 2025A03J3960).
It was also supported by the Knut and Alice Wallenberg Foundation and the Swedish Foundation for Strategic Research. 
It was also partially supported by the MINERVA project, funded by the European Research Council (ERC) under the Horizon 2022 research and innovation program of the European Union (Grant agreement No. 101044629).} 
\thanks{Corresponding author: Apostolos I. Rikos.}
}
\maketitle

% ========================================
%
% Abstract
%
% ========================================
\begin{abstract} 
We present \acronymbf, a novel distributed, finite-time algorithm that guarantees max--consensus in directed networks with unreliable communication links experiencing packet drops. 
Unlike existing methods, \acronymbf ensures all nodes compute the exact maximum state under arbitrary packet loss patterns. 
It incorporates a fully distributed termination mechanism, enabling nodes to autonomously determine whether convergence has occurred. 
Our algorithm leverages narrowband error-free feedback channels to acknowledge successful (single-bit) transmissions with minimal communication overhead. 
We analyze our algorithm's operation, and we provide a convergence proof establishing explicit bounds on the required time steps. 
We validate its correctness in a wireless sensor network for environmental monitoring, and finally, we compare against existing approaches highlighting our algorithm's operational advantages. 
\end{abstract}


% ========================================
%
% Introduction
%
% ========================================
\section{Introduction}
\label{sec:introduction}

% \todo{ADD BELOW IN REFERENCES} 

% https://www.tandfonline.com/doi/epdf/10.1080/00207721.2019.1585998?src=getftr&utm_source=sciencedirect_contenthosting&getft_integrator=sciencedirect_contenthosting

% Max-consensus of multi-agent systems in random networks✩ Jianing Yang a , Liqi Zhou a , Bohui Wang b , Yuanshi Zheng

% \todo{ALSO ADD REFERENCES FROM LATTER ABOVE} 

\IEEEPARstart{T}{he} relentless surge of interconnectivity has led to a growing need for effective control and coordination within networks of multiple agents, such as groups of sensors \cite{2005:XiaoBoydLall} and mobile autonomous agents \cite{2004:Murray}.
A problem of particular interest in distributed control and coordination is the \textit{consensus} problem \cite{2020:Yang_Thomas_Survey_Consensus, 1974_DeGroot_118_121, 2005:Olshevsky_Tsitsiklis}. 
The {consensus} problem is to develop distributed algorithms that can be used by a group of agents in order to reach agreement to a common decision (or state). 
The agents engage in the process to reach consensus with different initial states/data and communicate locally with neighboring agents. 
The concept of max--consensus refers to the process of achieving the maximum state among a set of states held by different nodes in a network. 
Knowledge of the maximum value of a state is of importance for various applications such as leader election \cite{2020:Berenbrink_Leader_Kling}, clock and data synchronization \cite{2015:Ishii_ClockSynch, 2007:Gamba}, termination mechanisms \cite{2015:Cady_Hadjicostis_2664_2670}, distributed clustering \cite{2022:Rikos_kmeans_CDC}, fault detection \cite{2011:Shames_Johansson_Fault_Detection}, resource allocation \cite{2021:Luo_Weisong_Survey_Resource, 2023:Grammenos_Themis_1880_1894, 2023:Rikos_CPU_TCNS}, and dynamic quantization mechanisms~\cite{L-CSS2025:Oliva}.

% ========================================
% Related Work
% ========================================
\subsection{Related Work}

% \todo{check if there are more works + enhance in description and make our work stand out}

The problem of max--consensus has received significant attention in the literature. 
Asynchronous operation has been a focal point in several studies \cite{2012:Iutzeler_Jakubowicz, 2016:Giannini_Rizzo_MaxConsensus, 2023:Rikos_CPU_TCNS}. 
In \cite{2012:Iutzeler_Jakubowicz} randomized algorithms with asynchronous operations were explored, demonstrating convergence within finite time with high probability. 
Furthermore, in \cite{2016:Giannini_Rizzo_MaxConsensus}, algorithms operating asynchronously with transmissions affected by time delays were presented. 
Additionally, in \cite{2023:Rikos_CPU_TCNS}, asynchronous coordination algorithms with quantized communication that incorporate max-- and min--consensus operations were proposed. 
Conversely, synchronous operation took precedence in other works \cite{2008:Cortes, 2020:Mangal_Salapaka_DistrStopping_Delays, 2009:Monajemi_Jorg_MaxConsensus}.  
For example in \cite{2008:Cortes} nodes operated synchronously and achieved max--consensus within finite time over static directed graphs. 
In \cite{2020:Mangal_Salapaka_DistrStopping_Delays} scenarios involving synchronous algorithms with transmission delays were examined to detect convergence of the underlying protocol. 
Furthermore, in \cite{2009:Monajemi_Jorg_MaxConsensus} max-plus algebra was employed to analyze the convergence rate of synchronous max--consensus algorithms for time-invariant networks. 
The dynamics of time-varying networks were analyzed in \cite{2006:Tahbaz_Jadbabaie_MaxConsensus, 2012:Shi_Johansson_MaxConsensus}.  
In \cite{2006:Tahbaz_Jadbabaie_MaxConsensus}, a family of consensus algorithms was introduced for time-varying networks, demonstrating connections to Bellman-Ford and mean hitting time iterations under certain modifications. 
In related work, \cite{2012:Shi_Johansson_MaxConsensus} focused on max--consensus over time-varying and state-dependent graphs, establishing necessary and sufficient conditions for convergence. 
Unreliable networks were the main focus in \cite{2019:Nowzari_Rabbat_MaxConsensus, 2010:Nejad_Raisch_MaxConsensus, 2019:Muniraju_Spanias_MaxConsensus}. 
Specifically, \cite{2019:Muniraju_Spanias_MaxConsensus} explored cases where edges were absent with non-zero probabilities independently across edges and time, \cite{2019:Nowzari_Rabbat_MaxConsensus} calculated tighter convergence bounds for two algorithms achieving max--consensus in asynchronous networks, and \cite{2010:Nejad_Raisch_MaxConsensus} considered scenarios where messages were dropped based on non-zero probabilities. 
Prior approaches on the max--consensus problem considered also a diversity of alternative perspectives. 
For instance, in \cite{2017:Hendrickx_Heemels_MaxConsensus}, the focus was on scenarios where the size of the communication network underwent dynamic changes. 
In \cite{2023:Franceschelli_Guia_MaxOConsensus}, two distinct algorithms were proposed to address the tracking of time-varying maximum states within networks, addressing scenarios where the maximum state evolves over time. 
The work \cite{2020:Venkategowda_Werner_1839_1843} focused on the domain of privacy preservation during maximum state calculations. 
Furthermore, the effects of additive communication noise were studied in \cite{2019:Muniraju_Spanias_MaxConsensus, 2016:Zhang_Spanias_9089_9098}, where max-plus algebra was employed as a tool for analyzing the ergodic process.

% ========================================
% Motivation
% ========================================
\subsection{Motivation}

Despite the extensive body of literature on max--consensus, existing approaches predominantly provide {probabilistic} convergence guarantees, often relying on specific statistical assumptions about network disturbances or link reliability.
Consequently, there is no absolute certainty that all nodes will {deterministically} compute the correct maximum within a finite number of steps under arbitrary packet loss patterns. 
In many real‑world scenarios (especially in safety‑critical, high‑stakes, or resource‑constrained applications) the absence of such guarantees is unacceptable. 
Even a small probability of error or unbounded convergence time can lead to suboptimal, delayed, or unsafe decisions; for example, in environmental monitoring, resource allocation, or fault detection in safety-critical systems. 
Deterministic convergence is therefore essential when correctness and timeliness are non‑negotiable.
Moreover, most prior works do not equip nodes with a fully {distributed} mechanism for detecting when convergence has been achieved.
Without such a mechanism, nodes risk terminating prematurely, compromising correctness, or continuing computation and communication unnecessarily, wasting both energy and bandwidth (this could be a particularly severe problem for various applications involving battery‑powered or large‑scale deployments).
These shortcomings leave a significant open challenge: to design a distributed max--consensus protocol that not only operates reliably over networks with unreliable communication links, but also delivers {deterministic, finite‑time} convergence together with accurate, fully local convergence detection. 
% Such a protocol would bridge the gap between theoretical guarantees and real‑world applicability, enabling trustworthy deployment in critical networked systems. \todo{Themi is this last sentence ok?}

% ========================================
% Main Contributions
% ========================================
\subsection{Main Contributions} 

Motivated by the need for {deterministic} max--consensus under arbitrary packet loss patterns and the absence of a fully distributed convergence detection mechanism, we make the following key contributions in this paper:
\begin{list4}
\item We introduce \acronym, the first distributed algorithm that guarantees finite-time convergence to the exact maximum state in directed networks with unreliable communication links, {without} imposing restrictive probabilistic assumptions on packet drops (Algorithm~\ref{algorithm_CUL}).
\item We provide a built-in distributed termination detection mechanism by exploiting narrowband error-free feedback channels. 
Hence, \acronym enables each node to locally determine, in a fully distributed manner, when the global maximum has been reached, thus ensuring correct termination even under arbitrary link failures.
\item We prove that \acronym converges in finite time for any packet drop pattern satisfying weak connectivity conditions (i.e., the probability of a packet being dropped is less than $1$), and we derive an explicit probabilistic upper bound on the number of required iterations (Theorem~\ref{theorem_convergence}). 
% \item We demonstrate the effectiveness and applicability of \acronym in an environmental monitoring setting, where a sensor network must compute the maximum measured temperature despite unreliable links, highlighting both its accuracy and its efficiency (Section~\ref{sec:results}). 
\item We demonstrate the effectiveness and applicability of \acronym in an environmental monitoring setting, where a wireless sensor network deployed in an outdoor environment must compute the maximum measured temperature in finite time despite packet drops. 
In this scenario, each node maintains the maximum of its local measurements and periodically runs \acronym to obtain the network-wide maximum under possible energy constraints (Section~\ref{sec:results}). 
% The required $1$-bit feedback is carried by existing acknowledgment mechanisms in low-power protocols (e.g., IEEE 802.15.4, LoRaWAN, BLE Mesh), incurring negligible additional bandwidth or energy overhead (Section~\ref{sec:results}).
\end{list4}
\textbf{Paper Organization.} 
In Section~\ref{sec:notation} we provide the notation, information regarding the underlying network, node operation, and we present the problem formulation. 
In Section~\ref{algorithm_max_cons} we present our proposed algorithm and in Section~\ref{sec:CONValgorithm_max_packet} we analyze its operation establishing its finite time convergence. 
In Section~\ref{sec:results} we illustrate our algorithms' performance on an application and compare it with existing works. 
Section~\ref{sec:future} concludes the paper. 

% ========================================
%
% Preliminaries
%
% ========================================
\section{Preliminaries}
\label{sec:notation}

% \subsection{Notation}\label{graph_notation}

\textbf{Notation.} The sets of real, rational, integer, and natural numbers are denoted by $ \mathbb{R}, \mathbb{Q}, \mathbb{Z}$, and $\mathbb{N}$, respectively. 
The symbol $\mathbb{Z}_+$ denotes the set of nonnegative integers and the symbol $\mathbb{N}_0$ denotes the set of natural numbers that also includes zero.  
The $\div$ operation represents the division of one number by another, returning the whole number quotient. 
For example, $\alpha \div \beta$ denotes the division of the dividend $\alpha$ by the divisor $\beta$ resulting in the largest whole number quotient. 
The $\%$ operation represents the modulo or remainder when one number is divided by another. 
For example, in the expression $\alpha \ \% \ \beta$ the dividend $\alpha$ is divided by the divisor $\beta$, and the modulo operation returns the remainder. 
The $\max$ operation, also known as the maximum operation, returns the greater of two given values. 
It can be represented as $\max(\alpha, \beta)$ where $\alpha$ and $\beta$ are the values being compared.
The expected value of a random variable $x$ is denoted as $\mathrm{E}[x]$, and represents the theoretical mean or average outcome that would be obtained if the random variable were repeatedly observed.
For any real number $a \in \mathbb{R}$, the floor $\lfloor a \rfloor$ denotes the greatest integer less than or equal to $a$ while the ceiling $\lceil a \rceil$ denotes the least integer greater than or equal to $a$. 

%Consider a network of $n$ ($n \geq 2$) agents communicating only with their immediate neighbors. 
\textbf{Communication Network.} 
The communication network consists of $n \geq 2$ nodes communicating only with their immediate neighbors and can be captured by a \textit{static} directed graph (digraph). 
A static digraph is defined as $\mathcal{G}_d = (\mathcal{V}, \mathcal{E})$ with $n = |\mathcal{V}|$ nodes and $m = |\mathcal{E}|$ edges. 
The set of nodes is defined as $\mathcal{V} =  \{v_1, v_2, \dots, v_n\}$, and the set of edges (self-edges excluded) is defined as $\mathcal{E} \subseteq \mathcal{V} \times \mathcal{V} \setminus \{ (v_j, v_j) \ | \ v_j \in \mathcal{V} \}$. 
A directed edge from node $v_i$ to node $v_j$ is denoted by $(v_j, v_i) \in \mathcal{E}$, and captures the fact that node $v_j$ can receive information from node $v_i$ (but not the other way around). 
The subset of nodes that can directly transmit information to node $v_j$ is called the set of in-neighbors of $v_j$ and is represented by $\mathcal{N}_j^- = \{ v_i \in \mathcal{V} \; | \; (v_j,v_i)\in \mathcal{E}\}$. 
The subset of nodes that can directly receive information from node $v_j$ is called the set of out-neighbors of $v_j$ and is represented by $\mathcal{N}_j^+ = \{ v_l \in \mathcal{V} \; | \; (v_l,v_j)\in \mathcal{E}\}$. 
The cardinality of $\mathcal{N}_j^-$ is called the \textit{in-degree} of $v_j$ and is denoted by $\mathcal{D}_j^-$ (i.e., $\mathcal{D}_j^- = | \mathcal{N}_j^- |$). 
The cardinality of $\mathcal{N}_j^+$ is called the \textit{out-degree} of $v_j$ and is denoted by $\mathcal{D}_j^+$ (i.e., $\mathcal{D}_j^+ = | \mathcal{N}_j^+ |$). 
The diameter $D'$ of a digraph is defined as the longest shortest path between any two nodes $v_j, v_i \in \mathcal{V}$ in the network.

\subsection{Packet Dropping Links and Feedback Channels}\label{model_packet_drops}

In many real-world scenarios, distributed systems and networks often operate in environments characterized by unreliable communication links \cite{2020:Assran_Rabbat, 2018:BOOK}. 
Understanding and addressing the challenges posed by such unreliable networks is of utmost importance for ensuring the robustness and efficiency of distributed algorithms. 
% For the development of the results in this paper, we make the following assumption. 

% \todo{[It is not an assumption, it is your system model]} 

% \begin{assum}\label{assum_packet_drops} 
\textbf{Modeling Packet Dropping Links.} 
In our setting, each communication link in the distributed network is unreliable (i.e., it is subject to packet drops or losses). 
Packet drops can occur due to network congestion, channel impairments, or other factors. 
The probability of a packet being dropped (or lost) at link $(v_j, v_i)$ is denoted by $q_{ji}$ (where $0 \leq q_{ji} < 1$), and is assumed to be independent and identically distributed across different links and time instances. 
% \end{assum}
The occurrence of packet drops is modeled by utilizing a Bernoulli random variable. 
Specifically, we define $x_k(j,i)$ as an indicator variable ($x_k(j,i) = 1$ indicates a successful transmission from node $v_i$ to node $v_j$ at time step $k$). 
The probability distribution of $x_k(j,i)$ is given by
\begin{equation}\label{dropsmodel}
\Pr\{ x_k(j,i)=m \} = \left\{ \begin{array}{ll}
         q_{ji}, & \mbox{if $m = 0$,}\\
         1 - q_{ji}, & \mbox{if $m = 1$,}\end{array} \right.
\end{equation}
where $q_{ji}$ represents the probability of a packet being dropped at the link $(v_j, v_i)$, and $1-q_{ji}$ represents the probability of a successful transmission.

\textbf{Modeling Feedback Channels.}
%In various message transmission protocols such as Transmission Control Protocol (TCP) and High-Level Data Link protocol, a mechanism called Automatic Repeat reQuest (ARQ) is utilized \cite{Krouk:2011, Garcia_Widjaja:2000}. 
%ARQ serves as an error control protocol, ensuring reliable packet transmissions over unreliable communication channels. 
%The operation of ARQ relies on a feedback signal transmitted from the data receiver to the data transmitter. 
%The feedback signal indicates whether a packet has been successfully received or not. 
%This feedback signal is transmitted via a narrowband feedback channel, for which we make the following assumption. 
In many communication protocols, such as Transmission Control Protocol and High-Level Data Link Control, reliable data delivery over unreliable links is achieved through the {Automatic Repeat reQuest} (ARQ) mechanism \cite{Krouk:2011, Garcia_Widjaja:2000}, an error control protocol, ensuring reliable packet transmissions over unreliable communication channels. 
ARQ employs a lightweight feedback signal, sent from the receiver back to the transmitter, to indicate whether a transmitted packet was successfully received. 
This feedback is typically carried over a narrowband channel, requires only a single bit per acknowledgment, and is therefore well suited for low-rate, low-power control signaling.
Motivated by this, we make the following assumption about the availability of such a feedback channel in our network model.
\begin{assum}\label{assum_feedback_channel}
For each directed link $(v_j, v_i) \in \mathcal{E}$ from node $v_i$ to node $v_j$, there exists a narrowband error-free feedback channel from node $v_j$ to node $v_i$. 
\end{assum}

The feedback channel introduced in Assumption~\ref{assum_feedback_channel} usually comprises messages of $1$ bit (i.e., it is not used for data transmission but solely to acknowledge whether the packet was successfully received or not). 
In practice, the assumption of a narrowband error-free feedback channel can be realized in a variety of ways, particularly in low-power wireless sensor networks (WSNs). 
Many widely used low-power protocols, such as IEEE 802.15.4, LoRaWAN, and BLE Mesh, already incorporate acknowledgment mechanisms at the MAC layer to ARQ. Furthermore, narrowband channels can be implemented using sub-GHz radios (e.g., 433/868/915 MHz) that offer better propagation and interference resilience compared to broadband channels, particularly in outdoor monitoring applications.
For this reason, it is reasonable to assume that the feedback channel is error-free. 
This is also a standard assumption in the literature~\cite{2019GunduzTWC,2021:JSAC_HARQ,2022:Makridis_Themis_Hadj,Makridis2023CDC,CHARALAMBOUS2024}.
%\TC{As we will see in our algorithm development later,} such a feedback mechanism is necessary for acknowledging whether a packet has been received or not. 
%In our setting, the feedback mechanism is also utilized to notify each node in the network whether the operation of the algorithm could be terminated or continue to be executed. 
% it is also important for communicating the state of each node with respect to whether it needs another round of transmissions or not, thus notifying each node in the network whether the operation of the algorithm could be terminated or continue to be executed.  
In our algorithm, this feedback not only confirms packet reception but also enables each node to learn, in a fully distributed manner, whether to terminate or continue execution. 

\subsection{Node Variables}\label{subs_feedback_ch}

At time step $k$, each node $v_j \in \mathcal{V}$ maintains the following variables: 
$x_j[k]$, the node's state; $x^{\rm old}_j$, for checking whether the node's state was updated during a set of max--consensus operations; $\text{dflag}_j$, a flag that is used to notify nodes in the network whether the operation of the algorithm is terminated or needs to be continued; $R_{ji}$, used from node $v_j$ to mark its incoming links from which it received information at least once. 
% Furthermore, for the operation of every node in the network we make the following assumption. 

% ========================================
%
% Preliminaries
%
% ========================================
\subsection{Problem Formulation}
\label{sec:probForm}

Consider a digraph $\mathcal{G}_d = (\mathcal{V}, \mathcal{E})$.
Each node $v_j \in \mathcal{V}$ has an initial state $x_j[0] \in \mathbb{R}$ and $q_m$ is the maximum state among the initial states defined as  
\begin{equation}\label{real_max}
q_m = \max_{v_j \in \mathcal{V}}\{x_j[0]\} . 
\end{equation}
Assume that each communication link in $\mathcal{G}_d$ is subject to packet drops (see Section~\ref{model_packet_drops}). 

In this paper, we aim to develop a distributed algorithm that allows nodes to (i) calculate $q_m$ in a finite number of time steps and, (ii) terminate their operation {simultaneously} afterwards. 
More specifically, there exists a time step $k_0$ so that for every $v_j \in \mathcal{V}$ we have 
\begin{subequations}
\begin{align}
& x_j[k_0] = q_m,\label{realmax_cons_1} \\
& \text{nodes terminate operation at time step} \  k_0 .\label{realmax_cons_2} 
\end{align}
\end{subequations} 

% \noindent
% \textbf{P1.} \todo{FIX} \\ 
% When the digraph $\mathcal{G}_d$ is static and strongly connected, the algorithm allows the nodes to obtain, after a finite number of steps, a quantized state $q^s$ which is equal to the ceiling or the floor of the actual average $q$ of the initial states in \eqref{real_av}. 
% Specifically, we require that there exists $k_0$ so that for every $v_j \in \mathcal{V}$ we have
% \begin{equation}\label{alpha_q_no_oscill}
% ( q^s_j[k] = \lfloor q \rfloor \ \ \text{for} \ \ k \geq k_0 ) \ \ \text{or} \ \ ( q^s_j[k] = \lceil q \rceil \ \ \text{for} \ \ k \geq k_0).
% \end{equation}

% ========================================
%
% Main Results
%
% ========================================
\section{Main Results}\label{algorithm_max_cons}

% In this section, we present a max--consensus algorithm that can operate in networks with unreliable communication links. 
% Our algorithm is called \acronym (see Algorithm~\ref{algorithm_CUL}) and solves the problem described in Section~\ref{probForm}.
In this section, we present our \acronym algorithm that solves the problem described in Section~\ref{sec:probForm}.
Before presenting our algorithm, we consider the following assumptions which are important for the development of the results in this paper.

\begin{assum}\label{str_conn}
We assume that the digraph $\mathcal{G}_d$ is \textit{strongly connected}. 
\end{assum} 

\begin{assum}\label{diam_known}
We assume that every node $v_j$ knows an upper bound $D'$ of the diameter $D \leq D'$ of $\mathcal{G}_d$. 
\end{assum} 

Assumption~\ref{str_conn} means that for each pair of nodes $v_j, v_i \in \mathcal{V}$, $v_j \neq v_i$, there exists a directed \textit{path} from $v_i$ to $v_j$, i.e., information from every node can reach any other node in the network.  
Assumption~\ref{diam_known} is important for coordinating the actions of nodes in the network, and notifying them whether convergence has been achieved. 
Computing an upper bound on the network diameter can be easily performed in a distributed way within a finite number of steps (can be done as an initialization step) with quantized communications~\cite{RIKOS:2023networksize}.
%Note that the max--consensus has been extensively used in the literature as part of other processes \cite{2022:Liu_Wang}. 
%Therefore, the need to know the network diameter $D'$ is for completing the current procedure and moving on to the next process. 

The algorithm \acronym is based on two phases. 
\textit{Phase--1} is executed in Iteration-step~$1$ and lasts $D'$ time steps. 
At the beginning of this phase, each node $v_j$ marks each incoming link with the value one. 
Then, for $D'$ time steps, each node $v_j$ executes the max--consensus protocol. 
During the execution of max--consensus, 
if node $v_j$ received at least one message from in-neighbor $v_i\in\mathcal{N}^{-}_j$, it marks the link $(v_j, v_i)$ with the value zero. 
\textit{Phase--2} is executed in Iteration-step~$2$ and also lasts $D'$ time steps. 
At the beginning of this phase, each node $v_j$ checks the following:  
\begin{list4}
    \item[\emph{i)}] if during \textit{Phase--1}, there is at least one incoming link from which it did not receive any message (i.e., there is at least one incoming link marked with value one). 
    Then, it sets its $\text{dflag}_j$ equal to one.
    \item[\emph{ii)}] if during \textit{Phase--1}, its value was updated (i.e., increased), then it sets its $\text{dflag}_j$ equal to one. 
\end{list4} 
Then, for $D'$ time steps it transmits its $\text{dflag}_j$ to its in-neighbors. 
It receives the $\text{dflags}$ of its out-neighbors and updates its own $\text{dflag}_j$ to be equal to the maximum value between the stored and the received dflags. 
Finally, at the end of \textit{Phase--2} (i.e., after $D'$ time steps), if its $\text{dflag}_j$ is equal to zero, it terminates its operation. 
Otherwise, if its $\text{dflag}_j$ is equal to one, it goes to \textit{Phase--1} and repeats the operation.

The algorithm is described in detail in Algorithm~\ref{algorithm_CUL}. 

%\\ \noindent
%\textit{Phase--1}: Each node $v_j$ marks each incoming link with the value one. 
%Then, for $D'$ time steps, each node $v_j$ executes the max--consensus protocol. 
%During these $D'$ time steps, if it received at least one message from each incoming link, it marks this link with the value zero. 
%\\ \noindent
%\textit{Phase--2}: Each node checks: (i) if during \textit{Phase--1}, there is at least one incoming link from which it did not receive any message (i.e., there is at least one incoming link marked with value one), then it sets its $\text{dflag}_j$ equal to one, and (ii) if during \textit{Phase--1}, its value was updated (i.e., increased), then it sets its $\text{dflag}_j$ equal to one. 
%Then, for $D'$ time steps it transmits its $\text{dflag}_j$ to its in-neighbors. 
%It receives the $\text{dflags}$ of its out-neighbors and updates its own $\text{dflag}_j$ to be equal to the maximum value between the stored and the received flags. 
%Finally, at the end of \textit{Phase--2} (i.e., after $D'$ time steps), if its $\text{dflag}_j$ is equal to zero, it terminates its operation. 
%Otherwise, if its $\text{dflag}_j$ is equal to one, it goes to \textit{Phase--1} and repeats the operation.}

% \todo{I would distinguish Phase 1 and 2 in the algorithm}
\begin{varalgorithm}{1}
\caption{\acronym} 
\noindent \textbf{Input:} Digraph $\mathcal{G}_d = (\mathcal{V}, \mathcal{E})$ with $n=|\mathcal{V}|$ nodes and $m=|\mathcal{E}|$ edges. 
Each node $v_j \in \mathcal{V}$ has an initial state $x_j[0] \in \mathbb{R}$. \\ 
% \todo{Assumptions~\ref{assum_feedback_channel}, \ref{str_conn}, \ref{diam_known}, hold. [what this has to do with what each node does in the algo?]} \\
\textbf{Initialization:} Each node $v_j \in \mathcal{V}$ sets $x^{\rm old}_j = x_j[0]$, $\text{dflag}_j = 1$, and $R_{ji} = 1$ for every $v_i \in \mathcal{N}_j^-$. \\ 
\textbf{Iteration:} For $k= 0, 1, 2,\dots$, each node $v_j \in \mathcal{V}$, does the following:

\underline{\textit{Phase--1}:} if $k \div D' = 2\alpha$ (where $\alpha \in \mathbb{Z}_+$) then 
\begin{itemize}
\item if $k \ \% \ D'$ = 0, then if $\text{dflag}_j = 0$, then terminate operation; 
\item else: 
\begin{list4a}
\item[$1a)$] set $\text{dflag}_j = 0$; 
\item[$1b)$] broadcast $x_j[k]$ to every $v_l \in \mathcal{N}_j^+$; 
\item[$1c)$] receive $y_i[k]$ from every $v_i \in \mathcal{N}_j^-$, and set 
\begin{equation}\label{max_cons_upd}
x_j[k+1] = \max_{v_i \in \mathcal{N}_j^-}\{x_j[k] , w_{ji}[k] x_i[k] \} , 
\end{equation}
where $w_{ji}[k] = 1$ if node $v_j$ receives $y_{i}[k]$ from $v_i \in \mathcal{N}_j^-$ at iteration $k$ (otherwise $w_{ji}[k] = 0$); 
\item[$1d)$] if $w_{ji}[k] = 1$, for in-neighbor $v_i \in \mathcal{N}_j^-$, then it sets $R_{ji} = 0$; 
\end{list4a} 
\end{itemize}

\vspace{0.2cm} 

\underline{\textit{Phase--2}:} if $k \div D' = 2\alpha + 1$ (where $\alpha \in \mathbb{Z}_+$) then 
\begin{list4a} 
\item[$2a)$] if $R_{ji} = 1$ for at least one $v_i \in \mathcal{N}_j^-$ set $\text{dflag}_j = 1$; 
\item[$2b)$] if $x_j[k] > x^{\rm old}_j$ set $\text{dflag}_j = 1$; 
\item[$2c)$] broadcast $\text{dflag}_j$ to every in-neighbor $v_i \in \mathcal{N}_j^-$; 
\item[$2d)$] receive $\text{dflag}_l$ from every out-neighbor $v_l \in \mathcal{N}_j^+$ and set 
\begin{equation}\label{max_cons_upd_flag}
\text{dflag}_j = \max_{v_l \in \mathcal{N}_j^+}\{\text{dflag}_j, \text{dflag}_l\};  
\end{equation}
\item[$2e)$] set $x^{\rm old}_j = x_j[k]$, and $R_{ji} = 1$ for every $v_i \in \mathcal{N}_j^-$;
\item[$2f)$] set $k = k+1$ and go to \textit{Phase--1}.
\end{list4a}
\textbf{Output:} \eqref{realmax_cons_1}, \eqref{realmax_cons_2} hold for every $v_j \in \mathcal{V}$. 
\label{algorithm_CUL}
\end{varalgorithm}

We next provide an example to illustrate our algorithm's operation.

\begin{example}\label{Ex2}
Consider the digraph in Fig.~\ref{prob_example}, which consists the set of nodes $\mathcal{V} = \{ v_1, v_2, v_3 \}$, and the set of edges $\mathcal{E} = \{ (v_2, v_1) , (v_1, v_2),$ $(v_3, v_2) , (v_2, v_3), (v_3, v_1) \}$. 
Nodes $v_1$, $v_2$, $v_3$ have initial states $x_1[0] = 5$, $x_2[0] = 4$, $x_3[0] = 3$, respectively. 
Furthermore, the network diameter is $D'=2$, and Assumptions~\ref{assum_feedback_channel} -- \ref{diam_known}, hold. 
During the initialization of \acronym, each node $v_j$ sets $x^{\rm old}_j = x_j[0]$, $\text{dflag}_j = 1$, and $R_{ji} = 1$ for every $v_i \in \mathcal{N}_j^-$. 

\begin{figure}[t]
\begin{center}
\includegraphics[width=0.58\columnwidth]{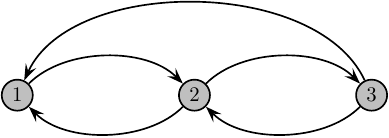}
\caption{Digraph in Example~\ref{Ex2}.}
\label{prob_example}
\end{center}
\end{figure}

During \textit{Phase--1}, at time step $k=0$ each node $v_j$ check its $\text{dflag}_j$. 
Since all dflags are equal to one, the operation is not terminated. 
Then, each node executes the max--consensus protocol for $D'$ time steps. 
Let us assume that during $k \in \{ 0, 1 \}$ time steps, link $(v_3, v_2)$ had two packet drops, while every other link had successful transmissions. 
This means that the states of nodes become: $x_1[2] = 5$, $x_2[2] = 5$, $x_3[2] = 3$. 
Furthermore, $R_{32} = 1$ and $R_{ji} = 0$, for every $(v_j, v_i) \in \mathcal{E} \setminus \{(v_3, v_2)\}$. 
At \textit{Phase--2}, at time step $k=2$, node $v_3$ sets $\text{dflag}_3 = 1$ because $R_{32} = 1$. 
Furthermore, node $v_2$ sets $\text{dflag}_2 = 1$ because $x_2[2] > x_2^{\rm old} = 4$). 
Then, for time steps $k= \{ 2, 3 \}$, nodes will transmit their dflags to their in-neighbors, and all dflags will become equal to one. 
Furthermore, each node $v_j$ will update its $x_j^{\rm old}$ value and its $R_{ji}$ value, for every $v_i \in \mathcal{N}_j^-$. 

At time step $k=4$, all dflags are equal to $1$. 
For this reason, every node will repeat \textit{Phase--1} {(i.e., the operation is not terminated).} 
% (i.e., the operation is not terminated). 
%It is important to note here that 
\acronym is not terminated, because during time steps $k \in \{ 0, 1 \}$ (i.e., during \textit{Phase--1}) (i) at least one link did not have any successful transmission (i.e., link ($v_3, v_2$)), and (ii) at least one node's state was increased (i.e., node $v_2$). 

During time steps $k \in \{ 4, 5 \}$ of \textit{Phase--1}, let us assume that every link had successful transmissions. 
% link $(v_1, v_3)$ had two packet drops, while every other link had successful transmissions. 
This means that the states of nodes become: $x_1[6] = 5$, $x_2[6] = 5$, $x_3[6] = 5$. 
Furthermore, 
% $R_{13} = 1$ and 
$R_{ji} = 0$ for every 
% $(v_j, v_i) \in \mathcal{E} \setminus \{(v_1, v_3)\}$. 
$(v_j, v_i) \in \mathcal{E}$. 
At \textit{Phase--2}, at time step $k=6$, 
% node $v_1$ sets $\text{dflag}_1 = 1$ because $R_{13} = 1$. 
% Furthermore, 
node $v_3$ sets $\text{dflag}_3 = 1$ because $x_3[6] > x_3^{\rm old}$ (note that $x_3^{\rm old} = 3$). 
Then, for time steps $k= \{ 6, 7 \}$, nodes will transmit their dflags to their in-neighbors, and all dflags will become equal to one. 
Furthermore, each node $v_j$ will update its $x_j^{\rm old}$ value and its $R_{ji}$ value, for every $v_i \in \mathcal{N}_j^-$. 

At time step $k=8$, all dflags are equal to one, and every node will repeat again \textit{Phase--1}. 
Note that \acronym is not terminated, because during time steps $k \in \{ 4, 5 \}$ (i.e., during \textit{Phase--1}) at least one node's state was increased (i.e., node $v_3$). 

% At time step $k=8$, every node will repeat \textit{Phase1} (since all dflags became equal to one from \textit{Phase2}, the operation is not terminated). 

During time steps $k \in \{ 8, 9 \}$, of \textit{Phase--1}, let us assume that every link had successful transmissions. 
This means that the states of nodes are: $x_1[10] = 5$, $x_2[10] = 5$, $x_3[10] = 5$. 
Furthermore, $R_{ji} = 0$, for every $(v_j, v_i) \in \mathcal{E}$. 
At \textit{Phase--2}, at time step $k=10$, we have that $\text{dflag}_j = 0$, for every $v_j \in \mathcal{V}$ (since for every $v_j$, we have $x_j[10] = x_j^{\rm old}$, and $R_{ji} = 0$ for every $v_i \in \mathcal{N}_j^-$). 
Then, for time steps $k= \{ 10, 11 \}$, nodes will transmit their dflags to their in-neighbors.
In this case, we have that all dflags will remain equal to \textit{zero}. 
Furthermore, each node $v_j$ will update its $x_j^{\rm old}$ value and its $R_{ji}$ value, for every $v_i \in \mathcal{N}_j^-$. 

Finally, at time step $k=12$ every node will check its dflag. 
Note that all dflags are equal to zero because during time steps $k \in \{ 8, 9 \}$ (i.e., during \textit{Phase--1}) (i) every link had at least one successful transmission, and (ii) the state of every node did not increase. 
For this reason, the operation of \acronym will be terminated. 
% the state of every node is equal to $5$ (which is the maximum value in the network). 
The states of the nodes during the algorithm's operation are shown in Fig.~\ref{plot3nodes_examp}. 

\begin{figure}[t]
\centering
\includegraphics[width=\columnwidth]{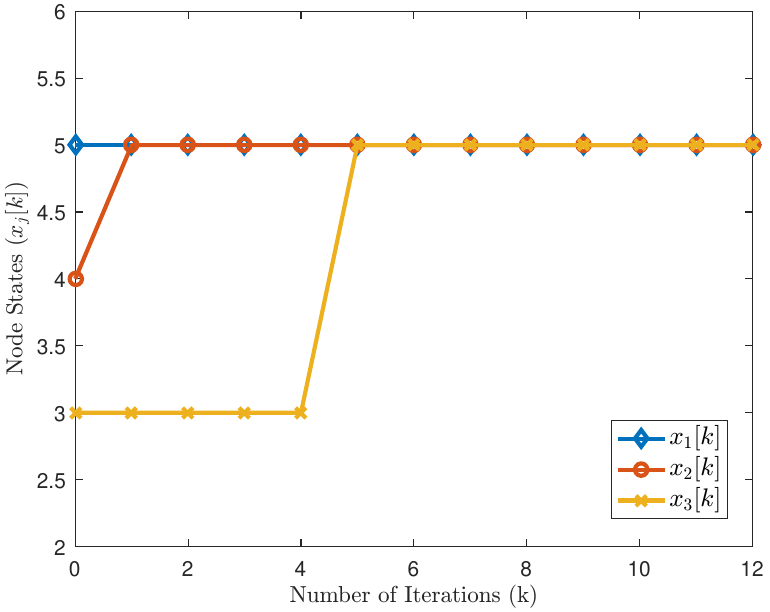}
\caption{Node states plotted against the number of iterations during execution of \acronym for the digraph shown in Fig.~\ref{prob_example}.}
\label{plot3nodes_examp}
\end{figure}
\end{example}

% ========================================
%
% Convergence Analysis
%
% ========================================
\section{Convergence Analysis}
\label{sec:CONValgorithm_max_packet}

{We now analyze the convergence of \acronym. 
We show that during the algorithm's operation there exists a time step $k_{\mathrm{terminal}}$ for which \eqref{realmax_cons_1}, \eqref{realmax_cons_2} hold for every node $v_j$ after a finite number of time steps. 
Then, we calculate an explicit upper bound on $k_{\mathrm{terminal}}$.} 

\begin{theorem}\label{theorem_convergence}
Consider a strongly connected digraph $\mathcal{G}_d = (\mathcal{V}, \mathcal{E})$ with $n = |\mathcal{V}|$ nodes and $m = |\mathcal{E}|$ edges. 
Each node $v_j \in \mathcal{V}$ has an initial state $x_j[0] \in \mathbb{R}$ and Assumptions~\ref{assum_feedback_channel}, \ref{str_conn}, and \ref{diam_known} hold.  
During the execution of \acronym, there exists a time step $k_{\mathrm{terminal}} \in \mathbb{N}$ for which \eqref{realmax_cons_1} and \eqref{realmax_cons_2} hold for every node $v_j \in\mathcal{V}$. 
This allows the algorithm to terminate in a distributed fashion. 
\end{theorem}

\begin{proof}
We present three parts.  
In \textit{Part--1}, we establish that our algorithm converges for the simple case where every transmission from every link is successful (i.e., in \eqref{dropsmodel} we have $ \Pr\{ x_k(j,i) = 1 \} = 1$ for every $(v_j,v_i) \in \mathcal{E}$). 
Also, in \textit{Part--1} we highlight the main differences from the standard max--consensus algorithm. 
In \textit{Part--2}, we show that our algorithm converges to the correct maximum for the case where one link has packet drops for $D'$ consecutive time steps only, while every other transmission from that particular link (excluding these $D'$ consecutive time steps) and every other link is successful.  
In \textit{Part--3}, we extend to the general case in which every link suffers from packet drops (i.e., from Section~\ref{model_packet_drops} we have $0 < q_{ji} < 1$ for $(v_j,v_i) \in \mathcal{E}$) and we show that our algorithm converges to the maximum.

\textit{Part--1}: 
During the execution of \acronym, each node performs max--consensus for time steps $k = 0, ..., D' - 1$. 
This ensures that by time step $D'$, the state of every node satisfies \eqref{realmax_cons_1} (i.e., $x_j[D] = q_m$). 
During time step $k = D'$, at least one node $v_j$ sets $\text{dflag}_j = 1$ (because its state was increased). 
Therefore, during time steps $k = D', ..., 2D'- 1$, every node $v_j$ sets $\text{dflag}_j = 1$ (see Iteration Steps $2c$ and $2d$). 
% Subsequently, during time steps $k = D', ..., 2D'- 1$, every node $v_j$ sets $\text{dflag}_j = 1$.
% The purpose of setting $\text{dflag}_j = 1$ for every node is because at least one node (e.g., $v_i$) updates its state during time steps $k = 0, ..., D'- 1$. 
% Therefore, at time step $k = D'$, node $v_i$ sets $\text{dflag}_i = 1$, and subsequently, during time steps $k = D', ..., 2D'- 1$, every node $v_j$ sets $\text{dflag}_j = 1$ (see Iteration Steps $2c$ and $2d$). 
At time step $k = 2D'$, every node $v_j$ has $\text{dflag}_j = 1$. 
For this reason, each node performs max--consensus for an additional $D'$ time steps (i.e., for time steps $k = 2D', ..., 3D'- 1$). 
Following this, during time steps $k = 3D', ..., 4D'- 1$, each node $v_j$ sets $\text{dflag}_j = 0$ (note that since by time step $D'$, the state of every node satisfies \eqref{realmax_cons_1}, then no node increased its state during time steps $k = 2D', ..., 3D'- 1$).
Finally, at time step $4D'$, each node $v_j$ checks its $\text{dflag}_j$. 
Since $\text{dflag}_j = 0$, then every node terminates its operation. 
Thus, for \textit{Part--1}, our algorithm converges to the maximum after $4D'$ time steps. 

\textit{Discussion of Part--1:}
Compared to the max--consensus algorithm (e.g., \cite{2008:Cortes}), the \acronym algorithm is initially executed for $2D'$ time steps. 
Then, it is executed for an additional number of $2D'$ time steps every time (i) a node updates its state, and/or (ii) a packet from a link is dropped for $D'$ consecutive time steps (as it will be shown in \textit{Part--2} below). 
To this regard, in \textit{Part--1} nodes terminated their operation at time step $4D'$ since no node updated its state and no packet was dropped during time steps $k = 2D', ..., 3D'- 1$. 

\textit{Part--2}: 
Let's consider the simple case in which link $(v_l, v_j)$ drops packets at time steps $2D', ..., 3D'- 1$, while every other transmission from every link in the network is successful. 
During the operation of \acronym, every node will execute max--consensus for time steps $k = 0, ..., D'-1$. 
At time step $D'$, the state of every node fulfills \eqref{realmax_cons_1}. 
During time step $k = D'$, at least one node $v_j$ sets $\text{dflag}_j = 1$ (because its state was increased). 
Therefore, during time steps $k = D', ..., 2D'- 1$, every node $v_j$ sets $\text{dflag}_j = 1$. 
For this reason, every node will execute max--consensus for an additional number of $D'$ time steps (i.e., for time steps $k = 2D', ..., 3D'-1$). 
Note however that at time steps $2D', ..., 3D'- 1$ every packet at link $(v_l, v_j)$ is dropped. 
This means that at time step $k = 3D'$, node $v_l$ sets $\text{dflag}_l = 1$, and every node $v_j \in \mathcal{V} \setminus \{v_l\}$ sets $\text{dflag}_j = 0$. 
Consequently, for time steps $k = 3D', ..., 4D'-1$ every node $v_j \in \mathcal{V}$ will set $\text{dflag}_j = 1$ (see Iteration Steps~$2c$, $2d$). 
Since $\text{dflag}_j = 1$ for every $v_j$, every node will execute max--consensus for time steps $k = 4D', ..., 5D'-1$. 
Then, for time steps $k = 5D', ..., 6D'-1$ every node $v_j$ will set $\text{dflag}_j = 0$ (because during time steps $k = 4D', ..., 5D'-1$, every transmission from every link was successful, and no node increased its state). 
At time step $6D'$, since $\text{dflag}_j = 0$ for every $v_j$, then every node will terminate its operation. 
Therefore, for \textit{Part--2}, our algorithm will converge to the maximum after $6D'$ time steps.  

\textit{Discussion of Part--2:}
In \textit{Part--2}, the presence of $D'$ consecutive packet drops at link $(v_l, v_j)$ during time steps $2D', ..., 3D'- 1$ resulted in \acronym being executed for an extra $2D'$ time steps compared to \textit{Part--1}. 
This additional execution was necessary to account for the packet drops that occurred during time steps $2D', ..., 3D'- 1$, even though max--consensus had already been reached at time step $D'$.  

\textit{Part--3}: 
{In this part, suppose that there does not exist any node $v_l\in \mathcal{N}$ that is not receiving any packet within $D'$ consecutive time steps from any of its in-neighbors $v_j\in\mathcal{N}^{-}$ (i.e., at least on of the in-neighboring nodes will send a packet within $D'$ steps). Then, the only condition for the node $v_l$ to set its $\text{dflag}_l = 1$ is if it changes its value. Notably, the union of all the graphs formed at each of the $D'$ time steps is a strongly connected graph, and as a result, if a node has a different value than the other nodes, a  $\text{dflag}$ will be set to one at the end of \textit{Phase--2}. 
Therefore, the nodes in the network should continue executing \acronym until they ensure that they calculate the maximum of their states.} 
\end{proof}

In Theorem~\ref{theorem_convergence} we showed that our algorithm converges deterministically. 
In the following proposition, we establish a probabilistic bound on the number of time steps needed for every node to calculate the maximum state in the network.

\begin{prop}
Let us define the maximum packet drop probability among all links in the network as
\(q_{\text{max}} = \max_{(v_j, v_i) \in \mathcal{E}}\{q_{ji}\}\), and an epoch as the number of \(D'\) time steps.
For two parameters \(\varepsilon_0,\varepsilon_1 > 0\), chosen arbitrarily small, let us define \(k_0\) as the minimum number of time steps such that every link \((v_j,v_i) \in \mathcal{E}\) performs at least one successful transmission with probability at least \(1-\varepsilon_1\).
For \(k_0\) it holds
\begin{align}\nonumber
    k_0 \ge \left\lceil
\frac{\log\!\Bigl(\,1-(1-\varepsilon_0)^{1/|\mathcal{E}|}\,\Bigr)}{D'\log(q_{\max})} 
\right\rceil D'.
\end{align}
Furthermore, let us define \(k_1\) as the number of epochs for which no link has \(D'\) consecutive packet drops with probability at least \(1-\varepsilon_2\).
For \(k_1\) it holds
\begin{align}\nonumber
    k_1 \geq \Bigl \lceil \frac{\log(\varepsilon_1)}{\log(p_1)} \Bigr \rceil ,
\end{align}
where \(p_1\) is the probability that at least one link suffers packet drops for \(D'\) consecutive time steps.
With probability at least \(1-\varepsilon\), where \(\varepsilon = 1 - (1-\varepsilon_1)^{D'}(1-\varepsilon_2)\), the termination step \(k_{\mathrm{terminal}}\) for \acronym is given by
\begin{align}\nonumber
    k_{\mathrm{terminal}} = 2D'(k_0 + k_1),
\end{align}
where \(D'\) is the diameter of the network. 
\end{prop}

\begin{proof}
In \textit{Part--3} of Theorem~\ref{theorem_convergence}, we showed that \acronym converges to the maximum state after a finite number of time steps in a deterministic fashion. 
However, it is possible to calculate an upper bound on the time steps required for {achieving} convergence with high probability. 

% Let us consider the maximum packet drop probability among all links in the network defined as $q_{\text{max}} = \max_{(v_j, v_i) \in \mathcal{E}}\{q_{ji}\}$. 
% Also, let us define an epoch as the number of $D'$ time steps.
We first calculate a lower bound of the minimum number of time steps $k_0$, (for which it also holds $k_0\% D'=0$), for which every link $(v_j,v_i) \in \mathcal{E}$ will have performed at least one successful transmission with probability $1-\varepsilon_0$, where $\varepsilon_0$ can be arbitrarily small. 
To achieve this, we solve the inequality 
\begin{equation}\label{desiredprob_onetrans}
    (1 - q_{\text{max}}^{k_0})^{|\mathcal{E}|} \geq 1 - \varepsilon_0 \;.
\end{equation}
%where $1- \varepsilon_0$ represents a probability almost equal to one. 
In \eqref{desiredprob_onetrans} we solve for $k_0$, divide by $D'$ (to find the number of epochs) and take the ceiling of the result (because time steps are discrete values), multiplying by $D'$ in order to find the minimum number of time steps $k_0$ that satisfies $k_0\% D'=0$. 
Thus, from \eqref{desiredprob_onetrans} we obtain 
% \begin{equation}\label{time_step_one_instance}
%     k_0 \geq \Bigl \lceil \frac{\log(|\mathcal{E}|/\varepsilon_0)}{D \log(1/q_{\text{max}})} \Bigr \rceil . 
% \end{equation} 
\begin{equation}\label{time_step_one_instance}
k_0 \ge \left\lceil
\frac{\log\!\Bigl(\,1-(1-\varepsilon_0)^{1/|\mathcal{E}|}\,\Bigr)}{D'\log(q_{\max})} 
\right\rceil D'.
\end{equation} 
% Eq.~\eqref{time_step_one_instance} represents the epoch $k_0$ at which every link will have performed at least one successful transmission with probability $(1 - \varepsilon_0)$ almost equal to one. 
Let us suppose that we execute \textit{Phase--1} of \acronym for $k_0$ time steps ($k_0$ fulfills \eqref{time_step_one_instance}). 
This means that each link has at least one successful transmission during the $k_0$ time steps with probability $(1 - \varepsilon_0)$. 
Therefore, by setting $\varepsilon_0$ small enough, we have that after $D'k_0$ time steps during \textit{Phase--1} of our algorithm each node fulfills \eqref{realmax_cons_1} with probability $(1 - \varepsilon_0)^{D'}$. 

Continuing the algorithm's operation, consider the maximum packet drop probability for $D'$ consecutive time steps defined as $q_{\text{max}}^{D'}$. 
The probability $p_1$ that at least one link will suffer packet drops for $D'$ consecutive time steps is 
% \begin{equation}\label{d_cons_packet_dr}
%     p_1 \leq \sum_{\lambda = 1}^{|\mathcal{E}|} \binom{|\mathcal{E}|}{\lambda} (q_{\text{max}}^{D})^{\lambda} (1-q_{\text{max}}^{D})^{|\mathcal{E}| - \lambda} ,
% \end{equation}
\begin{equation}\label{d_cons_packet_dr_alternative}
p_1 \leq 1 - (1 - q_{\text{max}}^{D'})^{|\mathcal{E}|},
\end{equation}
where $|\mathcal{E}|$ is the number of links in the network, $D'$ is the network diameter, and $q_{\text{max}}$ is the maximum packet drop probability among all links. 
% \AR{
% Note here that an alternative and equivalent bound is
% \begin{equation}\label{d_cons_packet_dr_alternative}
% p_1 \leq 1 - (1 - q_{\text{max}}^{D})^{|\mathcal{E}|},
% \end{equation}
% which follows directly from the complement rule and the independence of packet drops across links. 
% } 
% \todo{rephrase as previous}
% We now calculate a number of epochs $k_1$ for which no link has $D'$ consecutive packet drops. 
% To achieve this we solve the inequality 
We now calculate a lower bound of the minimum number of epochs $k_1$ 
% (for which it also holds $k_1\% D'=0$), 
for which no link $(v_j,v_i) \in \mathcal{E}$ has $D'$ consecutive packet drops with probability $1-\varepsilon_1$, where $\varepsilon_1$ can be arbitrarily small. 
Note that \(k_1\) counts epochs, since \(p_1\) in \eqref{d_cons_packet_dr_alternative} denotes the probability that at least one link experiences \(D'\) consecutive packet drops over one block of \(D'\) time steps (i.e., within one epoch). 
To achieve this, we solve the inequality 
\begin{equation}\label{prob_stop}
    1 - p_1^{k_1} \geq 1 - \varepsilon_1 . 
\end{equation} 
% In \eqref{prob_stop} we solve for $k_1$, divide by $D'$ (to find the number of epochs) and take the ceiling of the result (because time steps are discrete values), multiplying by $D'$ in order to find the minimum number of time steps $k_1$ that satisfies $k_1\% D'=0$. 
Therefore, from \eqref{prob_stop} we have

\begin{equation}\label{time_step_no_D_drops}
    k_1 \geq \Bigl \lceil \frac{\log(\varepsilon_1)}{\log(p_1)} \Bigr \rceil . 
\end{equation} 
In \eqref{time_step_no_D_drops} we have that after $k_1$ epochs (or equivalently $D'k_1$ time steps) no link will have $D'$ consecutive packet drops with probability at least $1 - \varepsilon_1$. 

% epochs of $D'$ time steps each, no link will have $D'$ consecutive packet drops with probability at least $1 - \varepsilon_1$. 

Let us now combine the results in \eqref{time_step_one_instance} and \eqref{time_step_no_D_drops}. 
From \eqref{time_step_one_instance} we have that after $2 D'k_0$ time steps (i.e., $D'k_0$ time steps during \textit{Phase--1} and $D'k_0$ time steps during \textit{Phase--2} of our algorithm) each node fulfills \eqref{realmax_cons_1} with probability $(1 - \varepsilon_0)^{D'}$. 
Additionally, from \eqref{time_step_no_D_drops} we have that after $2 D'k_1$ time steps no link will have $D'$ consecutive packet drops with probability at least $1 - \varepsilon_1$ (since we have $Dk_1$ time steps during \textit{Phase--1} and $Dk_1$ time steps during \textit{Phase--2} of our algorithm). 
Combining the above results, we have that after $2 D'k_0 + 2 D'k_1$ time steps during our algorithm each node fulfills \eqref{realmax_cons_1}, and \eqref{realmax_cons_2} with probability at least $(1 - \varepsilon_0)^{D'}(1 - \varepsilon_1)$. 
This concludes our proof. 

\end{proof}

% ========================================
%
% Application: Distributed Environmental Monitoring in Sensor Networks
%
% ========================================
\section{Application: Distributed Environmental Monitoring in WSNs}
\label{sec:results}

We now demonstrate the practical applicability of \acronym in a wireless sensor network (WSN) deployed for outdoor environmental monitoring \cite{2019:Carminati_Marcuccio_44_52, 2018:Lombardo_Elsayed_1214_1222}. 
In this setup, nodes monitor local temperature and cooperatively compute the global maximum temperature via \acronym. 
Real‑world deployments face unreliable communication due to interference, obstacles, and energy constraints, leading to packet drops.
Since nodes may be operating under resource constraints (e.g.,
nodes are often battery‑powered) it is critical that they reach an accurate consensus in finite time and terminate automatically to preserve energy.

In our setup, each node periodically measures and stores its local maximum temperature. 
Nodes then execute \acronym to compute the global maximum (for example, at the end of each day or every $15$ minutes in a greenhouse for timely climate control). 
Additionally, the narrowband error‑free $1$-bit feedback required by \acronym can be implemented using existing acknowledgment mechanisms in low‑power WSN protocols (e.g., IEEE 802.15.4, LoRaWAN, BLE Mesh).
Such ACKs occupy only a single bit and can be embedded in MAC‑layer control frames with negligible overhead (making the approach suitable for energy‑constrained sensor nodes while preserving deterministic convergence guarantees under packet drops).

In order to highlight the behavior and advantages of \acronym algorithm, we focus on the three following cases. 
In \textbf{Case~$1$}, we evaluate our algorithm on a network for environmental monitoring modeled as a random digraph of $20$ nodes. 
Each edge of the digraph was generated with probability $0.2$ and is $D'= 4$. 
The packet drop probability for each link $(v_j, v_i) \in \mathcal{E}$ is set to $q_{ji} = 0.9$. 
We plot the state $x_j[k]$ of each node $v_j$ and the aggregate sum of $\text{dflag}_j$ against the number of executions of \textit{Phase--1} and \textit{Phase--2} (note that \textit{Phase--1} and \textit{Phase--2} are executed sequentially, i.e., when \textit{Phase--1} is executed, \textit{Phase--2} is executed immediately afterward). 
In \textbf{Case~$2$}, we analyze how the network diameter affects the operation of our algorithm. 
For this reason, we assess the performance of our algorithm over environmental monitoring networks modeled as randomly generated digraphs of $50$ nodes with diameters $D'$ chosen from the set $\{ 3, 5, 7 \}$. 
Each edge of the digraphs was generated with probability $0.2$, and the packet drop probability for each link $(v_j, v_i) \in \mathcal{E}$ is set to $q_{ji} = 0.9$. 
We plot the distribution of the number of executions of \textit{Phase--1} and \textit{Phase--2} required for our algorithm to converge. 
The results are averaged over $20$ executions to ensure robustness. 
In Case~$3$, we explore the influence of packet drop probabilities on the convergence of \acronym. 
We focus on an environmental monitoring network modeled as a random digraph of $50$ nodes. 
Each edge of the digraph was generated with probability $0.2$ and diameter is $D'= 4$. 
We execute our algorithm for $100$ instances for packet drop probabilities set to $q_{ji} = 0.9, 0.93, 0.96$, and $0.99$ for each $(v_j, v_i) \in \mathcal{E}$.
We plot the distribution of the number of executions of \textit{Phase--1} and \textit{Phase--2} required for our algorithm to converge. 

\begin{figure}[t]
\centering
\includegraphics[width=\columnwidth]{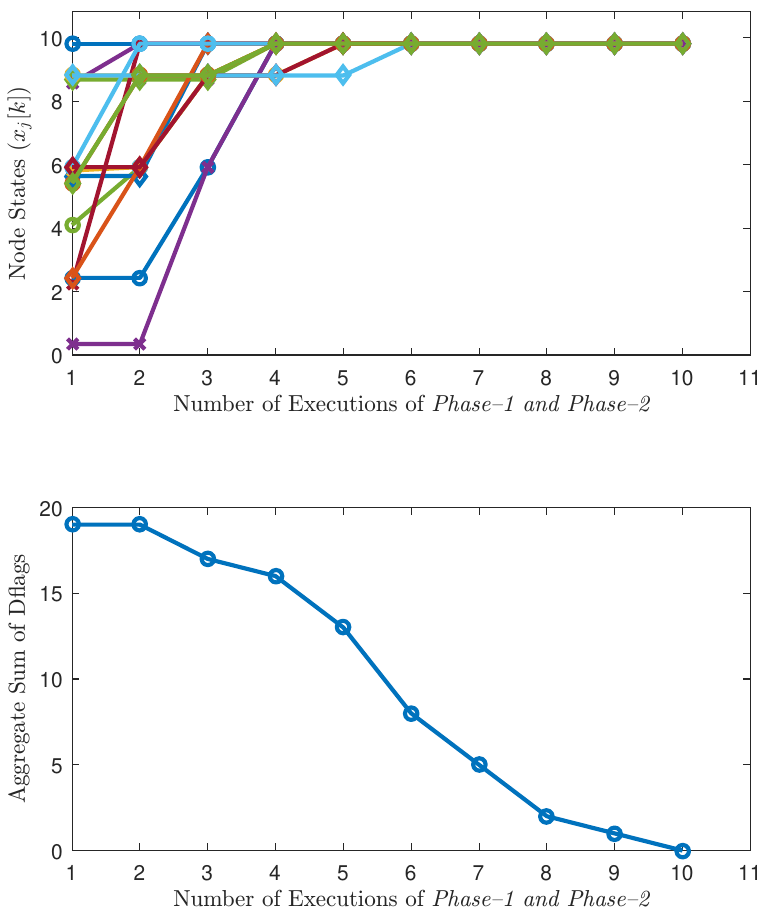}
\caption{Execution of \acronym over a random digraph of $20$ nodes with upper bound on diameter $D'=4$.} 
\label{case1_20nodes_D4_pack09}
\end{figure}

\textbf{Case~$1$.} 
{In Fig.~\ref{case1_20nodes_D4_pack09},} 
% we execute \acronym on a random digraph comprising $20$ nodes. 
% Each edge of the digraph was generated with probability $0.2$ and diameter is $D'= 4$.  
the top figure shows that each node successfully calculates the maximum value $q_m = 9.81$ after $10$ executions of \textit{Phase--1} and \textit{Phase--2}, and then terminates its operation. 
This convergence is also reflected in the bottom figure, where $\text{dflag}_j = 0$ for every node $v_j$ after $10$ executions.
It is interesting to note that all nodes are able to determine the maximum state in the network after $6$ executions of \textit{Phase--1} and \textit{Phase--2}. 
However, the presence of potential packet drops in the network extends the execution of our algorithm. 
Specifically, after $6$ executions, the aggregate sum of $\text{dflag}_j$ for each node $v_j$ is equal to $7$. 
This means that for $7$ nodes, at least one incoming link had $D'$ consecutive packet drops during \textit{Phase--1} (the probability of $D'$ consecutive packet drops is equal to $0.9^4$). 
As a result, an additional $4$ instances of \textit{Phase--1} and \textit{Phase--2} are executed in our algorithm (until the $\text{dflag}_j$ for each node $v_j$ becomes equal to $0$).

\begin{figure}[t]
\centering
\includegraphics[width=0.85\columnwidth]{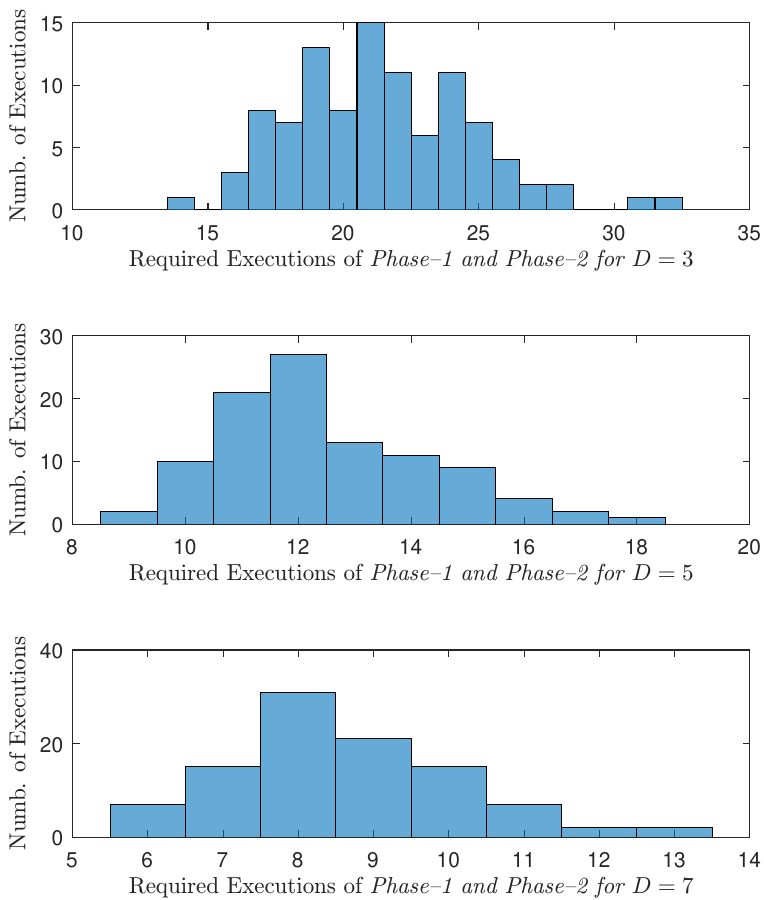}
\caption{Required executions of \textit{Phase--1} and \textit{Phase--2} for \acronym to converge, for random digraphs with diameters $D'=3, 5, 7$.}
\label{case2_50nodes_D357}
\end{figure}

\textbf{Case~$2$.} 
In Fig.~\ref{case2_50nodes_D357}, 
% we execute \acronym over random digraphs of $50$ nodes with diameters $D'= 3, 5, 7$. 
% Each edge was generated with probability $0.2$, and the packet drop probability for each link $(v_j, v_i) \in \mathcal{E}$ is set to $q_{ji} = 0.9$. 
for $D'= 3$ we can see that our algorithm converges after $17 - 23$ executions of \textit{Phase--1} and \textit{Phase--2} on average. 
However, it is important to notice that increasing the diameter to $D'= 5$, results in a decrease of the required executions of \textit{Phase--1} and \textit{Phase--2} for convergence. 
Specifically, for $D'= 5$ our algorithm converges after $11 - 13$ executions of \textit{Phase--1} and \textit{Phase--2} on average. 
This is mainly because the probability of $D'$ consecutive packet drops during \textit{Phase--1} for $D'= 5$ decreases to $0.9^5$ compared to $0.9^3$ for $D'= 3$. 
Finally, increasing the diameter, to $D'= 7$, further decreases the required executions of \textit{Phase--1} and \textit{Phase--2} for convergence. 
This occurs because the probability of $D'$ consecutive packet drops during \textit{Phase--1} further decreases to $0.9^7$ for $D'= 7$. 

\begin{figure}[t]
\centering
\includegraphics[width=0.85\columnwidth]{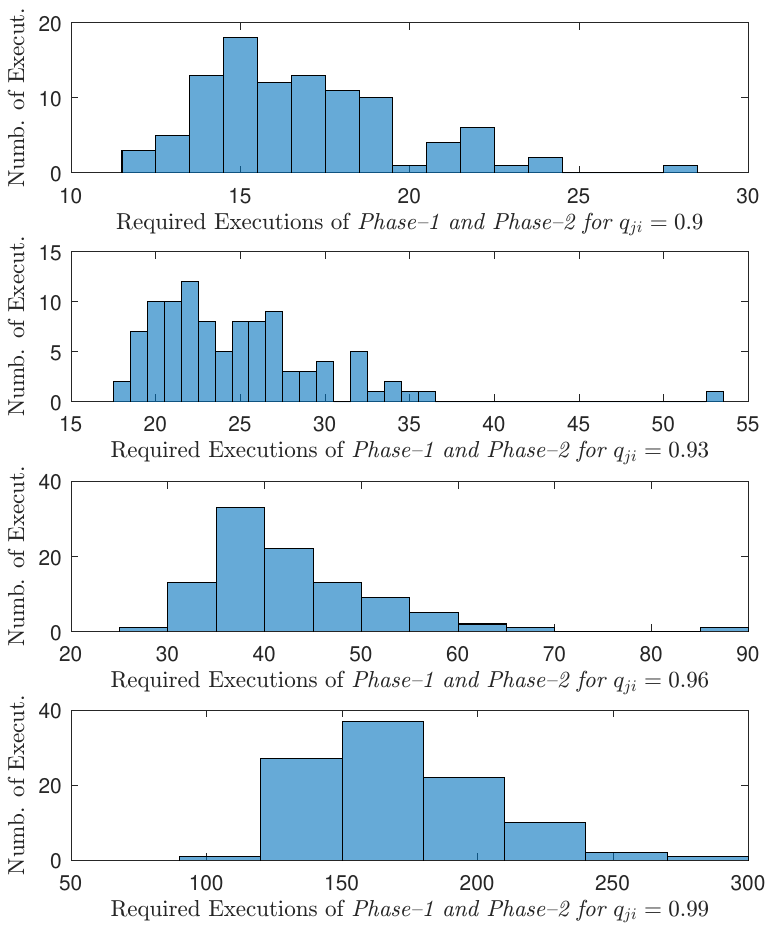}
\caption{Required executions of \textit{Phase--1} and \textit{Phase--2} for \acronym to converge, for $q_{ji} = 0.9, 0.93, 0.96$, and $0.99$ for each $(v_j, v_i) \in \mathcal{E}$, and diameter $D'= 4$.} 
\label{case3_50nodes_D4_V4}
\end{figure}

\textbf{Case~$3$.} 
{In Fig.~\ref{case3_50nodes_D4_V4},} 
% we execute \acronym on a random digraph comprising $50$ nodes. 
% Each edge of the digraph was generated with probability $0.2$ and diameter is $D'= 4$.  
% We present $100$ executions for the cases where each link $(v_j, v_i) \in \mathcal{E}$ has packet dropping probability equal to $q_{ji} = 0.9, 0.93, 0.96$, and $0.99$. 
for $q_{ji} = 0.9$ for each link $(v_j, v_i) \in \mathcal{E}$, we have that \textit{Phase--1} and \textit{Phase--2} are executed on average for $14 - 17$ instances for our algorithm to converge. 
For $q_{ji} = 0.93$, the number of executions of \textit{Phase--1} and \textit{Phase--2} slightly increases to $20 - 25$ instances. 
This is mainly because the probability of $D'$ consecutive packet drops during \textit{Phase--1} for $q_{ji} = 0.93$ increases to $0.93^4$ compared to $0.9^4$ for $q_{ji} = 0.9$. 
For $q_{ji} = 0.96$ the number of executions of \textit{Phase--1} further increases to $35 - 45$ (because the probability of $D'$ consecutive packet drops during \textit{Phase--1} further increases to $0.96^4$). 
Finally, for $q_{ji} = 0.99$ for each link $(v_j, v_i) \in \mathcal{E}$, the number of executions of \textit{Phase--1} and \textit{Phase--2} increases to $130 - 180$ instances (because $D'$ consecutive packet drops during \textit{Phase--1} occur with probability $0.99^4$). 

\begin{remark} 
In our analysis, we observed that as we increase the diameter $D'$, the number of executions of \textit{Phase--1} and \textit{Phase--2} decreases. 
Thus, increasing $D'$ improves the algorithm's convergence. 
However, let us note that increasing $D'$ also affects the total number of time steps required for convergence of our algorithm. 
As a result, although increasing $D'$ results in \textit{Phase--1} and \textit{Phase--2} being executed fewer times, each individual execution of \textit{Phase--1} and \textit{Phase--2} lasts longer (i.e., because \textit{Phase--1} and \textit{Phase--2} are executed for $D'+ D'= 2D'$ time steps). 
This trade-off between the number of executions of \textit{Phase--1} and \textit{Phase--2} and the duration of the execution of \acronym, highlights the importance of carefully considering the choice of the upper bound on the diameter $D'$ in applications of where convergence speed and communication overhead are critical factors.
\end{remark}

\begin{remark}
    Let us note here that another interesting application of our algorithm is for the case where nodes aim to calculate the average observed temperature in a finite time frame, with the added objective of terminating their operations upon calculation completion. 
    Calculation of the average temperature is achieved via average consensus algorithms that operate over networks consisted of unreliable communication links (e.g., see \cite{hadjicostis2015robust} and references therein). 
    Our proposed algorithm is able to operate in parallel with a robust average consensus algorithm. 
    This guarantees that nodes will be able to terminate their operation in a finite time frame and the average calculated temperature will be utilized for subsequent
    decision-making processes. 
\end{remark}

\subsection*{Comparison with Existing Literature} 

% \todo{2 PLOTS 
% \\
% ONE EVERYBODY CONVERGE BUT THEY DONT STOP TRANSMISSIONS
% \\
% SECOND THEY DONT KNOW DISTRIBUTION OF PACKET LOSSES, THEY HAVE UPPER BOUND ON THE NUMBER OF STEPS, NAD THEY DO NOT CONVERGE BECAUSE THEY ARE NOT ENOUGH}

% \begin{figure*}[t]
%     \centering
%     \begin{minipage}[b]{0.32\textwidth}
%         \centering
%         \includegraphics[width=\textwidth]{Figures/Ours_V2.pdf}
%         \label{fig:plot1}
%     \end{minipage}
%     \hfill
%     \begin{minipage}[b]{0.32\textwidth}
%         \centering
%         \includegraphics[width=\textwidth]{Figures/Cortes_V4.pdf}
%         \label{fig:plot2}
%     \end{minipage}
%     \hfill
%     \begin{minipage}[b]{0.32\textwidth}
%         \centering
%         \includegraphics[width=\textwidth]{Figures/Jakubo_V4.pdf}
%         \label{fig:plot3}
%     \end{minipage}
%     \caption{Comparison between \acronym (in $A-1$, $A-2$), \cite{2008:Cortes} (in $B-1$, $B-2$), and \cite{2012:Iutzeler_Jakubowicz} (in $C-1$, $C-2$), averaged over $20$ randomly generated strongly connected digraphs of $100$ nodes each.} 
%     \label{fig_compar_plots}
% \end{figure*} 

\begin{figure}[t]
\centering
\includegraphics[width=\columnwidth]{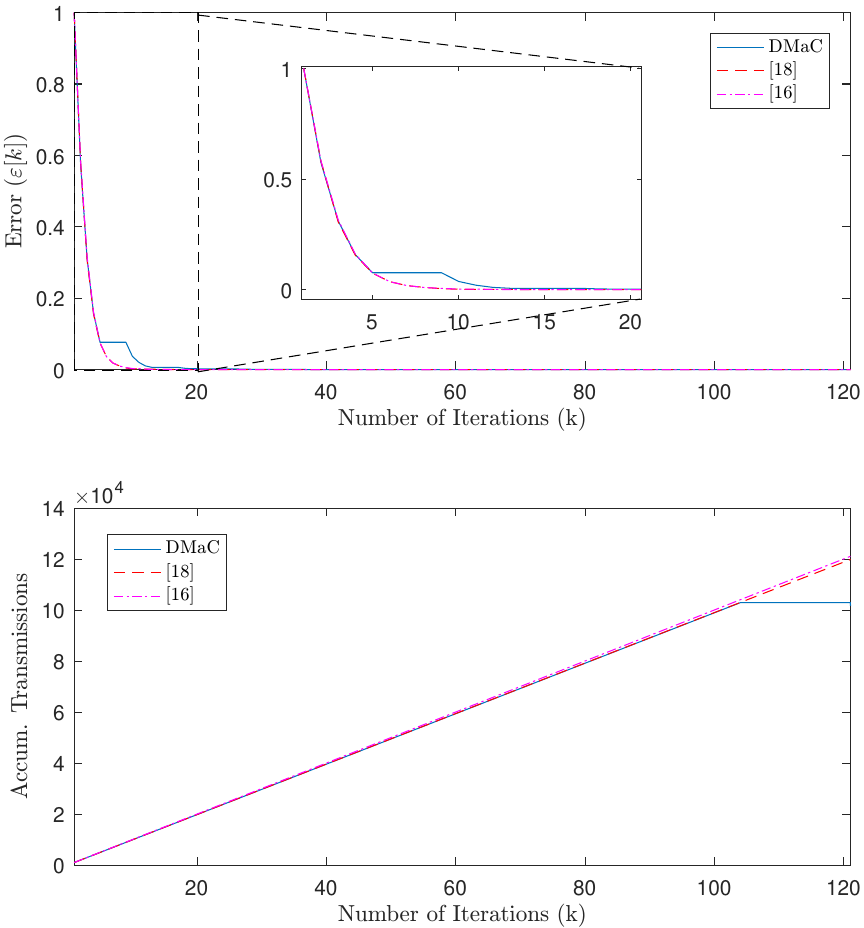}
\caption{Comparison between \acronym, \cite{2008:Cortes}, and \cite{2012:Iutzeler_Jakubowicz}, averaged over $20$ randomly generated strongly connected digraphs of $100$ nodes each.} 
\label{fig_compar_plots}
\end{figure}

We now compare the operation of \acronym against algorithms \cite{2008:Cortes, 2012:Iutzeler_Jakubowicz} from the literature. 
Our comparisons are shown in Fig.~\ref{fig_compar_plots}. 
During the operation of \acronym, and \cite{2008:Cortes, 2012:Iutzeler_Jakubowicz}, all parameters are consistent with \textbf{Case~$1$} above with the only differences being that our network is modeled as a random digraph of $100$ nodes, each edge of the digraphs was generated with probability $0.1$, and the network diameter is $D'= 6$. 
For \acronym, \cite{2012:Iutzeler_Jakubowicz}, and \cite{2008:Cortes} we plot 
\begin{itemize} 
    \item the error $\varepsilon[k]$ defined as
\begin{equation}
    \varepsilon[k] = \frac{\sum_{j=1}^n (x_j[k] - q_m)^2}{\sum_{j=1}^n (x_j[0] - q_m)^2}  , 
\end{equation}
where $q_m$ is defined in \eqref{real_max}, against the number of iterations, and 
    \item the total accumulated number of transmissions performed against the number of iterations. 
\end{itemize}
Note here that our results are averaged over $20$ randomly generated strongly connected digraphs of $100$ nodes each. 

In Fig.~\ref{fig_compar_plots} we have that during the operation of \acronym every node's state fulfills \eqref{realmax_cons_1} after $25$ iterations. 
Additionally, during \cite{2008:Cortes} and \cite{2012:Iutzeler_Jakubowicz} we have that every node's state fulfills \eqref{realmax_cons_1} after $10$ iterations. 
Additionally, we can see that during \acronym each node ceases transmissions and terminates its operation after $105$ iterations. 
This means that nodes fulfill \eqref{realmax_cons_2} after $105$ iterations. 
However, during \cite{2008:Cortes} and \cite{2012:Iutzeler_Jakubowicz} for $k \geq 10$, even though the state of every node fulfills \eqref{realmax_cons_1}, nodes continue performing transmissions towards their out-neighbors. 
Therefore, the operation of \cite{2008:Cortes}, and \cite{2012:Iutzeler_Jakubowicz} does not enable nodes to fulfill \eqref{realmax_cons_2}, as they continue performing transmissions unnecessarily, wasting energy and bandwidth resources. 
The above distinction makes \acronym the first algorithm in the literature that converges deterministically, enabling nodes to terminate their operation and cease transmissions after convergence and thus, improve resource efficiency which is necessary for resource‑constrained applications.

% ========================================
%
% Conclusions and Future Directions
%
% ========================================
\section{Conclusions}\label{sec:future}

% \subsection{Conclusions}
We introduced the \acronym algorithm for achieving max--consensus in finite time over networks with unreliable communication links. 
We showed that our algorithm enables nodes to calculate the maximum of their states deterministically by relying on narrowband error-free feedback channels. 
It also allows nodes to identify whether convergence has been achieved and terminate the process. 
We analyzed the algorithm's operation and established its convergence. 
Finally, we presented a motivating application in WSNs for environmental monitoring, and we compared our algorithm's operation against existing approaches from the literature, highlighting its unique advantages. 

% \subsection{Future Directions}

Future research will focus on broadening the resilience of our algorithm. 
First, we aim to extend the algorithm to operate under dynamic network conditions (e.g., open multi-agent systems and time-varying topologies), while preserving deterministic convergence guarantees. 
Another critical direction is enhancing robustness against adversarial attacks and Byzantine faults (where some nodes may behave maliciously or disseminate incorrect information). 
This will involve integrating fault-tolerant consensus techniques with our feedback-based convergence detection. 
Finally, we aim to focus on the case in which feedback channels themselves are unreliable and subject to packet drops. 

% ========================================
%
% References
%
% ========================================
%\section*{References}
%\def\refname{\vadjust{\vspace*{-1em}}} %Please don't do this in a real paper.
\bibliographystyle{IEEEtran}
\bibliography{bibliografia_consensus}

\end{document}